\documentclass[aps,prb,twocolumn,amsmath,amssymb,nofootinbib,tighten,showpacs,floatfix]{revtex4} 

\usepackage{graphicx}
\usepackage{dcolumn} 
\usepackage{bm}      

\begin{document}

\title{
Density of states of disordered Dirac particles: 
Infinitely many operators with negative scaling dimensions
and freezing transitions
      }

\author{Shinsei Ryu}
\affiliation{Department of Applied Physics,
             University of Tokyo,  
             7-3-1 Hongo Bunkyo-ku,
             Tokyo 113-8656,
             Japan}
                           
\author{Christopher Mudry}
\affiliation{Paul Scherrer Institute,
             CH-5232 Villigen PSI,
             Switzerland\\
             and Yukawa Institute for Theoretical Physics,
             Kyoto University,
             Kyoto 606-8502,
             Japan}  

\author{Akira Furusaki}
\affiliation{Yukawa Institute for Theoretical Physics,
             Kyoto University,
             Kyoto 606-8502,
             Japan}

\date{\today}

\begin{abstract} 
The global density of states (GDOS) close to the band center 
$\varepsilon=0$
for a particle hopping on a square lattice and
subjected to disorder that preserves the bipartite symmetry of the lattice
is computed using field theoretical methods.
The GDOS diverges like
$
|\varepsilon|^{-1}\exp
\left(
-c|\ln\varepsilon|^{\kappa}
\right)
$
with $\kappa=2/3$ in agreement with a prediction by
Motrunich \textit{et al.}\cite{Motrunich02}
and in disagreement with
an older prediction by Gade\cite{Gade93} ($\kappa=1/2$).

\end{abstract}


\maketitle

The problem of two-dimensional ($2D$) disordered Dirac particles,
particles with a relativistic dispersion
subjected to several types of quenched randomness,
has attracted a lot of interest since 1994.\cite{Ludwig94}
Indeed, it is related to a broad class of models 
such as the $2D$ random phase $XY$ model, 
dirty $d$-wave superconductors, or the Chalker-Coddington model for
the plateau transition in the integer quantum Hall effect,
in which disorder plays an essential role.
In spite of its relative simplicity, it can exhibit a rich variety of 
phenomena which appears to be specific to critical behavior 
induced by disorder. Multifractality is one example thereof.

Multifractal scaling implies
that the scaling regime in the vicinity of criticality 
is encoded by an \textit{infinite} set of independent relevant 
scaling exponents, the multifractal spectrum in short. 
On general grounds, the multifractal spectrum
must fulfill the condition of monotonicity.
Implementing this condition 
is a challenge to the common lore of \textit{local} 
field theoretical techniques. 

We consider a problem of Anderson localization defined on a $2D$
lattice where a particle is hopping on a square lattice 
in the background of $\pi$-flux phase
and subjected to weak uncorrelated bond-disorder 
that preserves the sublattice symmetry 
and time-reversal invariance.\cite{Hatsugai97}
The trademark of this model, the HWK model, is the
sublattice symmetry present for any given realization of the disorder.
Numerical studies offer strong evidences that typical wavefunctions 
at the band center are neither localized nor delocalized 
but are multifractal\cite{Hatsugai97} and that the density of states (DOS)
is singular\cite{Ryu02}. This disorder induced critical behavior
at the band center is caused by the sublattice symmetry.
After taking the continuum limit, the HWK model reduces to
two flavors of Dirac particles subjected to 
an imaginary Abelian random vector potential 
and a complex valued random mass.

It is known that monotonicity of the multifractal spectrum
for the zero-modes of one flavor of $2D$ Dirac particles subjected to 
a white-noise random vector potential
is implemented in a rather dramatic fashion, namely by the termination
of the multifractal spectrum.\cite{Chamon96}
In analogy to thermodynamics,
one can recast the termination of the multifractal spectrum as
a freezing transition.\cite{Castillo97}
{}From a field theoretical perspective,
the termination of the multifractal spectrum in this model
is closely related to the existence of an infinite set of 
operators with negative scaling dimensions
whose renormalization group (RG) flows are governed by 
a non-linear diffusion equation, the Kolmogorov-Petrovsky-Piscounov 
(KPP) equation.\cite{Mudry96,Carpentier00}
In this article, we sketch how, within field theory,
the global (i.e., self-averaging) density of states (GDOS) in the HWK model
is governed by the \textit{same} KPP equation
and diverges at the band center $\varepsilon=0$ as
$
|\varepsilon|^{-1}\exp
\left(
-c|\ln\varepsilon|^{\kappa}
\right)
$
with $\kappa=2/3$ in agreement with a prediction by
Motrunich \textit{et al.}\cite{Motrunich02}
and in disagreement with
an older prediction by Gade\cite{Gade93} ($\kappa=1/2$).

We start from the continuum limit of the HWK model.
The Hamiltonian is given by
($\sigma$'s are the Pauli matrices and the unit $2\times2$ matrix)
\begin{align}
H_{\hbox{\tiny{HWK}}}&=
\left(
\begin{array}{cc}
0 & D_{\hbox{\tiny{HWK}}} \\
D^{\dag}_{\hbox{\tiny{HWK}}} & 0 \\
\end{array}
\right),
\nonumber \\
D_{\hbox{\tiny{HWK}}}&=
i\sigma_{\mu}\partial_{\mu}
+i\sigma_{\mu} A_{\mu}
+i\sigma_{0} A_{0}
+\sigma_{3} A_{3},
\label{H_HWK}
\end{align}
where
$\boldsymbol{A}=(A_{1},A_{2})\in\mathbb{R}^2$
is the \textit{purely imaginary} random vector potential,
$A_0\in\mathbb{R}$
the \textit{purely imaginary} random scalar potential 
(chemical potential),
and
$A_3\in\mathbb{R}$ the random mass potential.
We assume that all potentials are white-noise Gaussian distributed with
vanishing means and with the variances specified by
$
\overline{
A^{\vphantom{*}}_\mu(x)A^{\vphantom{*}}_\nu(y)
         }
=
g^{\vphantom{2}}_{A}\delta_{\mu\nu}\delta(x-y)
$,
for $\mu,\nu=1,2$,
and
$
\overline{
A^{\vphantom{*}}_\mu(x)A^{\vphantom{*}}_\nu(y)
         }
=
g^{\vphantom{2}}_{M}\delta_{\mu\nu}\delta(x-y)
$,
for $\mu,\nu=0,3$.
Disorder averaging is denoted by an overline.
We employ the supersymmetric (SUSY) method and 
introduce four-component Grassmann fields
$\bar{\psi}$,
$\psi$,
and their bosonic partners
$\bar{\beta}$,
$\beta$
to represent the resolvent 
$\left(H_{\hbox{\tiny{HWK}}}-\varepsilon+i\eta\right)^{-1}$
of the HWK Hamiltonian in a given
realization of the disorder.
The advantage of the SUSY representation 
is that disorder averaging over the resolvent 
reduces to a Gaussian integration and induces an effective interacting 
theory with action $S_{\varepsilon-i\eta}$ for the fields 
$\bar{\psi}$,
$\psi$,
$\bar{\beta}$,
and
$\beta$.

Before characterizing the statistical distribution of the DOS,
we need to understand the action $S_{\varepsilon-i\eta=0}$.
When the random mass perturbation is switched off,
$g^{\vphantom{2}}_{M}=0$,
and at the band center,
$\varepsilon-i\eta=0$, 
the resulting effective interacting field theory 
$S_*$ defines a conformal field theory (CFT)
with $gl(2|2)$ current algebra symmetry 
for any value of $g^{\vphantom{*}}_{A}$.\cite{Guruswamy00}
It is also known that
$g^{\vphantom{*}}_M>0$ 
is exactly marginal whereas 
$g^{\vphantom{*}}_{A}$ 
is marginally relevant
with the non-perturbative beta function 
$
\beta^{\vphantom{*}}_{g^{\vphantom{*}}_{A}}=
(g^{\vphantom{2}}_{M}\zeta^{\vphantom{2}}_{})^{2}/(2\pi^{2})
$,
$
\zeta^{\vphantom{2}}_{}=
(1+ g^{\vphantom{2}}_{M}/\pi)^{-1}
$.\cite{Guruswamy00}
More precisely, switching on
a finite $g^{\vphantom{*}}_M>0$ demotes the CFT to the status of
a nearly conformal invariant theory (NCFT).\cite{Guruswamy00} 
Although conformal invariance is now broken, correlators for
the currents generating the $gl(2|2)$ algebra can nevertheless
be computed non-perturbatively in 
$g^{\vphantom{*}}_{A}$ and $g^{\vphantom{*}}_{M}$. 
Furthermore, a weak form of Wick theorem still holds in the NCFT.
This is all we will need to compute the GDOS.

To calculate the energy dependence of the GDOS, which we assume to
be identical to that of the disorder-average 
local density of states (LDOS)
$\nu_{\mathrm{av}}(\varepsilon)$,
one must reinstate a finite complex energy, i.e.,
a $GL(2|2)$ symmetry breaking perturbation 
$
\mathcal{A}_{N=1}=
\bar{\psi}\psi
+
\bar{\beta}\beta
$
is added to the action $S_{\varepsilon-i\eta=0}$
with the complex energy $\varepsilon-i\eta$ as a coupling constant.
The RG flow of the expectation value for
$\psi(x)\bar\psi(x)+\beta(x)\bar\beta(x)$
in the critical theory perturbed by a finite but small energy
under an infinitesimal rescaling 
$\mathfrak{a}\rightarrow \mathfrak{a}e^{dl}$
of the ultra-violet cutoff $\mathfrak{a}$
gives the Callan-Symanzik equation obeyed by
$\nu_{\mathrm{av}}(\varepsilon)$.

The energy perturbation $\mathcal{A}_{1}$
is one member of an infinite set of operators with negative
anomalous scaling dimensions.\cite{Mudry02}
Other members,
$\mathcal{A}_{N}=\frac{1}{N!}\mathcal({A}_{1})^{N}+\cdots$,
from this set are generated through the operator product expansion 
(OPE)\cite{Mudry02}
$
\mathcal{A}_{N}
\times
\mathcal{A}_{N^{\prime}}
=
\left(
\begin{array}{c}
N+N^{\prime}
\\
N
\end{array}
\right)
\mathcal{A}_{N+N^{\prime}}
+\cdots
$.
The anomalous scaling dimension 
$x^{\vphantom{*}}_N$
of 
$\mathcal{A}_{N}$
is\cite{Mudry02}
\begin{align}
x^{\vphantom{*}}_N&=
{\zeta^{\vphantom{2}}_{}}[1+\mathcal{O}(g^{\vphantom{2}}_{M})]N
-
\left[
\frac{g^{\vphantom{*}}_A}{\pi}{\zeta^{2}_{}}
+
\mathcal{O}(g^{\vphantom{2}}_{M})
\right]N^{2}.
\label{scaling_dim}
\end{align}
As the relevance of $\mathcal{A}_{N}$ grows with $N$ or $g^{\vphantom{*}}_A$,
all $\mathcal{A}_{N}$ 
must be taken into account on an equal footing
when performing a RG analysis of $\nu_{\mathrm{av}}(\varepsilon)$.

The perturbed action needed to perform a \textit{consistent} RG analysis 
of $\nu_{\mathrm{av}}(\varepsilon)$
is obtained by adding to 
$S_{\varepsilon-i\eta=0}$ the perturbations
$
-
\sum_{N=1}^{\infty}
Y^{\vphantom{*}}_{N}
\mathfrak{a}^{x_{N}-2}
\int d^2 r\,
\mathcal{A}_{N}(r)
$.
An expansion of the one-point correlation function
$\nu_{\mathrm{av}}(\varepsilon)$
up to second order in the ``fugacities'' $Y^{\ }_N$
shows that it obeys a Callan-Symanzik equation 
characterized by infinitely many 
one-loop $\beta$-functions 
$
\beta^{\vphantom{*}}_{Y^{\vphantom{*}}_N}\equiv
{dY^{\vphantom{*}}_{N}}/{dl}
$,
\begin{align}
\beta^{\vphantom{*}}_{Y^{\vphantom{*}}_N}=
(2-x^{\vphantom{*}}_N) Y^{\vphantom{*}}_N
+
\pi
\sum_{N^{\prime}=1}^{N-1}
\left(
\begin{array}{c}
N
\\
N^{\prime}
\end{array}
\right)
Y^{\vphantom{*}}_{N^{\prime}}
Y^{\vphantom{*}}_{N^{\vphantom{*}}-N^{\prime}}
\label{beta_fn}
\end{align}
with the initial condition
$
Y^{\vphantom{*}}_{N}(l=0)=
(\varepsilon-i\eta)/
\mathfrak{a}^{x^{\vphantom{*}}_{1}-2}
$,
together with an anomalous scaling \textit{operator} that couples the
expectation value of
$\psi(x)\bar\psi(x)+\beta(x)\bar\beta(x)$
to \textit{all} higher powers thereof.
This Callan-Symanzik equation is thus
an infinite set of coupled differential equations.

To treat all 
$\mathcal{A}_{N}$
on equal footing, we define the generating function
\begin{align}
\widetilde G(y,l)=
1
+
\frac{\pi}{2}
\sum_{n=1}^{\infty}
\frac{(-1)^n e^{-ny}}{n!}Y^{\vphantom{n}}_n(l).
\end{align}
The RG equation for this generating function reduces
to a non-linear diffusion equation
of the KPP-type,\cite{Carpentier00}
\begin{align}
\partial^{\vphantom{n}}_l \widetilde G=
\left(
{\zeta^{\vphantom{2}}_{}}
\partial^{\vphantom{n}}_y
+
\frac{g^{\vphantom{n}}_A{\zeta^{2}_{}}}{\pi}
\partial^{          2}_{y}
\right)\widetilde G
+2\widetilde G(\widetilde G-1)
\label{KPP_eqn}
\end{align}
with the initial condition
$
\widetilde G(y,0)=
\exp
\left(
-
\frac{\pi}{2}
\frac{
\varepsilon-i\eta
     }
     {
\mathfrak{a}^{x^{\vphantom{*}}_{1}-2}
     }
e^{-y}
\right)
$.
A detailed study of the asymptotic ($l\to\infty$)
solution of Eq.\ (\ref{KPP_eqn}) suggests
that the GDOS $\nu_{\mathrm{av}}(\varepsilon)$
is governed by the ``\textit{equivalent}''
Callan-Symanzik equation\cite{Mudry02}
\begin{align}
0&=
\left[
z^{\vphantom{*}}_A
\varepsilon
\frac{\partial}{\partial\varepsilon}
+
\beta^{\vphantom{*}}_{g^{\vphantom{*}}_A}
\frac{\partial}{\partial g^{\vphantom{*}}_A}
-
\left(
2-z^{\vphantom{*}}_A
\right)
\right]
\nu^{\vphantom{*}}_{\mathrm{av}}(\varepsilon).
\label{eq: Callan-Symanzik HWK typical DOS}
\end{align}
The infinite set of coupling constants $Y^{\vphantom{n}}_N$
together with the infinite dimensional anomalous scaling operator
have been replaced by a \textit{single} coupling constant 
$z^{\vphantom{*}}_A$, the dynamical scaling exponent, that 
relates a change of the \textit{typical energy scale} to 
a rescaling of length.
{}From the perspective of Eq.\ (\ref{KPP_eqn}), 
$z^{\vphantom{*}}_A$ is \textit{uniquely} defined as
the velocity of the asymptotic wave front solution.
It scales as $g^{\vphantom{*}}_{A}$ for $g^{\vphantom{*}}_{A}\zeta^2<2\pi$
whereas it scales as 
$\sqrt{g^{\vphantom{*}}_A}$ 
for 
$g^{\vphantom{*}}_{A}\zeta^2\geq2\pi$.
A freezing transition must necessarily take place at 
$g^{\vphantom{*}}_{A}\zeta^2=2\pi$
as $g^{\vphantom{*}}_{A}$ is marginally relevant.
Equation (\ref{eq: Callan-Symanzik HWK typical DOS})
can be solved. The most interesting case is that of weak disorder 
$g^{\vphantom{*}}_A(l=0)\zeta^2\ll1$ 
at the bare level.
Sufficiently far away from the band center, i.e., when the sublattice symmetry
is completely broken by the perturbing bare energy, the GDOS is flat. 
Upon approaching the band center,
the sublattice symmetry manifests itself by 
a power-law behavior of the GDOS
when the marginal relevance of
the Abelian random vector potential can be neglected.
At lower energies the relevance of 
$g^{\vphantom{*}}_{A}(l)\zeta^2\sim 1<2\pi$
cannot be neglected anymore
and the GDOS starts to increase in a Gade-like manner.
Arbitrary close to the band center
the relevance of $g^{\vphantom{*}}_A(l)\zeta^2\geq 2\pi$ 
dominates everything else in that the random vector potential
induces a strongly fluctuating LDOS (in a log-normal manner).
In the strong disorder regime $g^{\vphantom{*}}_A(l)\zeta^2\ge2\pi$,
$z^{\vphantom{*}}_A$ has undergone the freezing transition
that is responsible for zero-modes displaying localized behavior
instead of multifractality\cite{Chamon96}
and the GDOS diverging with the exponent $\kappa=2/3$ instead of 
$\kappa=1/2$\cite{Motrunich02}.

In summary, the HWK model is described in the continuum limit by a 
NCFT in which we identified infinitely many operators with 
negative anomalous scaling dimensions. These operators reflect the
broad distribution of the LDOS and enter in the field theoretical
computation of the GDOS through
a Callan-Symanzik equation made up of infinitely many differential
equations. From a detailed study of the RG equation obeyed
by a generating function for an infinite subset of relevant fugacities
we identified a scale dependent dynamical exponent that quantifies
the change in the typical energy scale due to a rescaling of length.
We then conjectured that the GDOS obeys an equivalent 
Callan-Symanzik equation made
of a finite number of differential equation and whose solution agrees with
the prediction from Motrunich \textit{et al.} 
for the GDOS in the HWK model.


\begin{thebibliography}{99}

\bibitem{Ludwig94} 
A.\ W.\ W.\ Ludwig, 
M.\ P.\ A.\ Fisher, R.\ Shankar, and G.\ Grinstein,
Phys.\ Rev.\ B \textbf{50}, 7526 (1994).

\bibitem{Hatsugai97}
Y.\ Hatsugai, 
X.-G.\ Wen, and M.\ Kohmoto,
Phys.\ Rev.\ B\ \textbf{56}, 1061 (1997).

\bibitem{Ryu02}
S.\ Ryu and Y.\ Hatsugai,
Phys.\ Rev.\ B\ \textbf{65}, 033301 (2002).

\bibitem{Chamon96}
C.\ C.\ Chamon, 
C.\ Mudry, and X.-G. Wen, 
Phys.\ Rev.\ Lett.\ \textbf{77}, 4194 (1996).

\bibitem{Castillo97}
H.\ E.\ Castillo, 
C.\ de C.\ Chamon, 
E.\ Fradkin, P.\ M.\ Goldbart, and C.\ Mudry,
Phys.\ Rev.\ B \textbf{56}, 10668 (1997).

\bibitem{Mudry96}
C. Mudry, 
C.\ Chamon, and X.-G. Wen, 
Nucl.\ Phys.\ \textbf{B466}, 383 (1996).

\bibitem{Carpentier00}
D.\ Carpentier and P.\ Le Doussal,
Nucl.\ Phys.\ \textbf{B588}, 565 (2000).
See also
B.\ Horovitz and P.\ Le Doussal,
Phys.\ Rev.\ B \textbf{65}, 125323 (2002).

\bibitem{Motrunich02}
O.\ Motrunich,
K.\ Damle, and D.\ A.\ Huse,
Phys.\ Rev. B \textbf{65}, 064206 (2002).

\bibitem{Gade93} 
R.\ Gade, Nucl.\ Phys.\ B \textbf{398}, 499 (1993);
R.\ Gade and F.\ Wegner, \textit{ibid.} \textbf{360}, 213 (1991).

\bibitem{Guruswamy00}
S.\ Guruswamy,
A.\ LeClair, and A.\ W.\ W.\ Ludwig,
Nucl.\ Phys.\  \textbf{B583}, 475 (2000).

\bibitem{Mudry02}
C.\ Mudry, S.\ Ryu, and A.\ Furusaki,
Phys.\ Rev.\ B \textbf{67}, 064202 (2003).

\end{thebibliography}
\end{document}